\documentclass[review,12pt]{elsarticle}

\newcommand{\g}{\,\vert\,}

\newcommand{\indpt}{\protect\mathpalette{\protect\independenT}{\perp}}
\def\independenT#1#2{\mathrel{\rlap{$#1#2$}\mkern2mu{#1#2}}}
\newcommand{\E}[1]{\mathbb{E}\left[#1\right]}
\newcommand{\R}{\mathbb{R}}
\newcommand{\EE}[2]{\mathbb{E}_{#1}\left[#2\right]}

\newcommand{\cN}{\mathcal{N}}
\newcommand{\cD}{\mathcal{D}}
\newcommand{\rom}[1]{\uppercase\expandafter{\romannumeral #1\relax}}

\newcommand{\dd}{\mathrm{d}}

\newcommand{\rmp}{\mathrm{p}}

\DeclareRobustCommand{\mb}[1]{\ensuremath{\boldsymbol{\mathbf{#1}}}}

\newcommand{\mbv}{\mathbf{v}}

\newcommand{\mbx}{\mathbf{x}}

\newcommand{\mbV}{\mathbf{V}}

\newcommand{\mbX}{\mathbf{X}}

\usepackage{amsmath}
\usepackage{caption}
\usepackage{subcaption}
\usepackage{hyperref}
\usepackage{xcolor}
\newtheorem{theorem}{Theorem}
\newtheorem{assumption}{Assumption}
\newtheorem{lemma}{Lemma}

\newproof{proof}{Proof}

\usepackage{amssymb}

\journal{arXiv}

\begin{document}

\begin{frontmatter}



\title{Adjusting for Indirectly Measured Confounding Using Large-Scale Propensity Score}


\author[1]{Linying Zhang}
\author[2]{Yixin Wang}
\author[3]{Martijn J. Schuemie}
\author[4,5]{David M.~Blei}
\author[1,6]{George Hripcsak}

\affiliation[1]{organization={Department of Biomedical Informatics, Columbia University Irving Medical Center},
            addressline={622 W. 168th Street, PH20}, 
            city={New York},
            postcode={10032}, 
            state={NY},
            country={USA}}
\affiliation[2]{organization={Department of Statistics, University of Michigan},
            addressline={1085 S University Ave}, 
            city={Ann Arbor},
            postcode={48109}, 
            state={MI},
            country={USA}}
\affiliation[3]{organization={Janssen Research and Development},
        addressline={1125 Trenton-Harbourton Road}, 
        city={Titusville},
        postcode={08560}, 
        state={NJ},
        country={USA}}          
\affiliation[4]{organization={Department of Statistics, Columbia University},
        addressline={1255 Amsterdam Ave}, 
            city={New York},
            postcode={10027}, 
            state={NY},
            country={USA}}
\affiliation[5]{organization={Department of Computer Science, Columbia University},
        addressline={500 West 120 Street, Room 450
MC0401}, 
            city={New York},
            postcode={10027}, 
            state={NY},
            country={USA}}
\affiliation[6]{organization={Medical Informatics Services, New York-Presbyterian Hospital},
            addressline={622 W. 168th Street, PH20}, 
            city={New York},
            postcode={10032}, 
            state={NY},
            country={USA}}           
\begin{abstract}
Confounding remains one of the major challenges to causal inference with observational data. This problem is paramount in medicine, where we would like to answer causal questions from large observational datasets like electronic health records (EHRs) and administrative claims. Modern medical data typically contain tens of thousands of covariates. Such a large set carries hope that many of the confounders are directly measured, and further hope that others are indirectly measured through their correlation with measured
covariates. How can we exploit these large sets of covariates for causal inference? To help answer this question, this paper examines the performance of the large-scale propensity score (LSPS) approach on causal analysis of medical data. We demonstrate that LSPS may adjust for indirectly measured confounders by including tens of thousands of covariates that may be correlated with them. We present conditions under which LSPS removes bias due to indirectly measured confounders, and we show that LSPS may avoid bias when inadvertently adjusting for variables (like colliders) that otherwise can induce bias. We demonstrate the performance of LSPS with both simulated medical data and real medical data.

\end{abstract}











\end{frontmatter}


\section{Introduction}
Causal inference in the setting of unmeasured confounding remains one of the major challenges in observational research. In medicine, electronic health records (EHRs) have become a popular data source for causal inference, where the goal is to estimate the causal effect of a treatment on a health outcome (e.g., the effect of blood-pressure medicine on the probability of a heart attack). EHRs typically contain tens of thousands of variables, including treatments, outcomes, and many other variables, such as patient demographics, diagnoses, and measurements.

Causal inference on these data is often carried out using propensity score adjustment \citep{rosenbaum1983}. Researchers first select confounders among the many observed variables, either manually (based on medical knowledge) or empirically. Then they estimate a propensity model using those selected variables and employ the model in a standard causal inference method that adjusts for the propensity score (the conditional probability of treatment). While this strategy is theoretically sound, in practice researchers may miss important confounders in the selection process, which leads to confounding bias, or may include variables that induce other types of bias (e.g., a ``collider" or a variable that induces ``M-bias").

In this paper, we study a closely related, but different, technique, known as large-scale propensity score (LSPS) adjustment \citep{tian2018}. LSPS fits an L1-regularized logistic regression with all pre-treatment covariates to estimate the propensity model. LSPS then uses standard causal inference methods, with the corresponding propensity scores, to estimate the causal effect. For example, LSPS might be used with matching \citep{rubin1973, rubin1973a, stuart2010} or subclassification \citep{rosenbaum1984}.

In contrast to the traditional approach of explicitly selecting confounders, LSPS is a ``kitchen-sink" approach that includes all of the covariates in the propensity model. While the L1-regularization might lead to a sparse propensity model, it is not designed to select the confounders in particular. Instead, it attempts to create the most accurate propensity model based on the available data, and LSPS diagnostics (described below) use covariate balance between treatment and control groups (i.e., that covariates are distributed similarly in the two groups) to assess whether \textit{all} covariates are in fact adjusted-for in the analysis regardless of their L1-regularization coefficient.

The discussion over how many covariates to include in a propensity model is an old one \citep{rubin2007, shrier2008, rubin2008, shrier2009, sjolander2009, pearl2009a, rubin2009} and considers--in the setting of imperfect information about variables--the tradeoff between including all measured confounders versus including variables that may increase bias and variance \citep{rubin1997, myers2011}. To address this issue, LSPS uses only pre-treatment covariates to avoid bias from mediators and simple colliders, and it uses diagnostics and domain knowledge to avoid variables highly correlated with the treatment but uncorrelated with outcomes (known as ``instruments"). Including such variables can increase the variance of the estimate \citep{rubin1997, myers2011, brookhart2006, austin2007} and can amplify bias \citep{pearl2010, pearl2011, pearl2013, wooldridge2016, steiner2016, ding2017}.

In medicine, empirical studies of the performance of LSPS have shown it to be superior to selecting confounders \citep{tian2018,ryan2013,weinstein2017,weinstein2020}. Consequently, LSPS has been used in a number of studies, both clinical \citep{lane2020,duke2017,morales2021,burn2019,wilcox2020,suchard2019,you2019,hripcsak2020,kim2020,you2020,vashisht2018} and methodological \citep{schuemie2021,schuemie2020,schuemie2020a,schuemie2018,schuemie2018a, schuemie2020b}.

Further, researchers have studied whether LSPS may also address indirectly measured confounders \citep{hripcsak2020, schuemie2020a, chen2020}. The hope behind these studies is that when we adjust for many covariates, we are likely to be implicitly adjusting for the confounders that are not directly measured but are correlated with existing covariates. Hripcsak et al.\cite{hripcsak2020} and Schuemie et al.\cite{schuemie2020a} used LSPS to estimate the causal effect of anti-hypertension drugs, adjusting for about 60,000 covariates. An important confounder, baseline blood pressure, was not contained in most of the data sources. In the one source that did contain blood pressure, adjusting for all the other covariates but no blood pressure resulted in (nearly) balancing blood pressure between propensity-score-stratified cohorts; the resulting causal inference was identical to the one obtained when including blood pressure in the propensity model. 

Based on this observation, Chen et al.\cite{chen2020} studied the effect of dropping large classes of variables from the LSPS analysis, using balance of the covariates between the treatment and control groups as a metric for successful adjustment (i.e., is every covariate in the propensity model balanced between the cohorts). If all the variables of one type were eliminated from the propensity model (e.g., medical diagnoses), then the inclusion of a large number of other variables (e.g., medications, procedures) resulted in the complete balancing of the missing variables. Even more striking, if all variables related to one medical area like cardiology were dropped from the model (e.g., all cardiology-related diagnoses, procedures, medications, etc.), then the rest of the covariates still balanced the dropped cardiology covariates. Yet if too few covariates were included, such as just demographics, then balance was not achieved on the other covariates. Based on these studies, LSPS appears to be adjusting for variables that are not included but correlated with the included covariates.

In this paper, we explore conditions under which LSPS can adjust for indirectly measured confounders. In particular, we provide some theoretical assumptions under which LSPS is robust to some indirectly measured confounders. They are based on the ``pinpointability" assumption used in Wang and Blei \cite{wang2019, wang2020}. A variable is pinpointed by others if it can be expressed as a deterministic function of them, though the function does not need to be known. In the context of causal inference from EHR data, we show that if confounders that are indirectly measured but can be pinpointed by the measured covariates, then LSPS implicitly adjusts for them. For example, if high blood pressure could be conceivably derived from the many other covariates (e.g., diagnoses, medicines, other measurements) then LSPS implicitly adjusts for high blood pressure even though it is not directly measured. 

From a theoretical perspective, pinpointability is a strong and idealized assumption. But in practice, several empirical observations showed that important confounders that are not directly measured often appear to undergo adjustment when LSPS is used. Therefore, there might be hope that some of the indirectly measured confounders are capturable by the existing covariates. We do not assert LSPS as a magical solution to unmeasured confounding—the assumption is strong—but as an attempt to better understand the empirical success of LSPS in adjusting for indirectly measured confounders. To explore this phenomenon, we use synthetic data to empirically study the sensitivity of LSPS to the degree to which pinpointability is violated. We find that under perfect pinpointability, adjusting for measured covariates removes the bias due to indirectly measured confounding. As the data deviates from pinpointability, adjusting for the measured covariates becomes less adequate.

Finally, we study real-world medical data to compare LSPS to a traditional propensity score method based on previously used manually selected confounders. We find that removing a known confounder has a bigger impact on a traditional propensity score method than on LSPS, presumably because it is indirectly measured. This finding suggests that including large-scale covariates with LSPS provides a better chance of correcting for confounders that are not directly measured.

The paper is organized as follows. Section 2 describes the LSPS algorithm, the pinpointability assumption, and the effect of pinpointability on M-structure colliders, instruments, and near-instruments. Section 3 studies the impact of violations of pinpointability on the fidelity of the estimated causal effects. Section 4 presents empirical studies comparing LSPS to classical propensity-score adjustment (with manually selected covariates), and methods that do not adjust. Section 5 compares LSPS to other approaches to adjusting for indirectly measured confounding and makes connection to other related work. Section 6 concludes the paper.

\section{The Large Scale Propensity Score Algorithm} \label{sec:methods}
In this section, we summarize the LSPS algorithm, describe an assumption under which LSPS will adjust for indirectly measured confounding and potentially mitigate the effect of adjusting for unwanted variables, and make some remarks on the assumption. 

\subsection{The LSPS algorithm}
We summarize the LSPS algorithm \citep{tian2018}, including the heuristics and diagnostics that normally surround it (e.g., Weinstein \cite{weinstein2017}). Consider a study where a very large number of covariates are available (e.g., over 10,000) and the problem of estimating the causal effect of a treatment. Rather than selecting confounding covariates and adjusting for them, LSPS adjusts for all of the available covariates. It uses only pre-treatment covariates to avoid adjusting for mediators and simple colliders (which induce bias), and it uses diagnostics and domain knowledge to avoid ``instruments," variables that are correlated with the treatment but do not affect the outcome. (Such variables increase the variance of the causal estimate.)

By design, LSPS includes all measured confounders. The hope is that in real-world data, such as in medicine, adjusting for all the other non-confounder variables would not impart bias, and empirical comparisons to traditional propensity approaches seem to bear that out \citep{tian2018,ryan2013,weinstein2017,weinstein2020}. The further hope is that by balancing on a large number of covariates, other indirectly measured factors would also become balanced, and this is what we address in Section \ref{subsec:pinpoint}.

The inputs to LSPS are observed pre-treatment covariates $\mbX$ and binary treatment $T$. The output is the estimated causal effect $\hat{\nu}$. LSPS works in the following steps.


\textbf{1. Remove ``instruments."} Remove covariates that are highly correlated with the treatment and are unlikely to be causally related to the outcome. Univariate correlation to treatment is checked numerically, and domain expertise is used to determine if the highly correlated variables are not causally related to the outcome; if the relationship is unclear, then the variable is not removed. Note these covariates are commonly called ``instruments," and used in instrumental variable analysis \citep{hernan2006}. LSPS, however, does not do instrumental variable analysis, and removes these variables to reduce downstream variance.

\textbf{2. Fit the propensity model and calculate propensity scores.} Given the remaining covariates, fit an L1-regularized logistic regression \citep{hastie2009} to estimate propensity scores $p(t \g \mbx)$. The regression is
    $$p(t \g \mbx) = \frac{1}{1+e^{-\mb\theta^\top\mbx}},$$
where $\mb\theta$ is the vector of the regression parameters. L1-regularized logistic regression minimizes
$$\mathcal{L}(\mb\theta) = \sum_{i=1}^N -t_i \log(p(t_i \g \mbx_i; \mb\theta))-(1-t_i) \log(1-p(t_i \g \mbx_i; \mb\theta)) + \lambda\sum_{j=1}^M | \theta_j |,$$
where $\lambda$ is the tuning parameter that controls the strength of the L1 penalty.
    
LSPS uses cross-validation to select the best regularization parameter $\lambda$. It then refits the regression model on the entire dataset with the selected regularization parameter. Finally, it uses the resulting model to extract the propensity scores for each datapoint.
    
 \textbf{3. Check the equipoise of the propensity model.} In this step, LSPS assesses whether the conditional distribution of assignment given by the propensity model is too certain, i.e., whether the treatment and control groups are too easily distinguishable. The reason is that a propensity model that gives assignments probabilities close to zero or one leads to high-variance estimates \citep{myers2011}, e.g., because it is difficult to match datapoints or create good subclasses.
 
 To assess this property of the propensity model, LSPS performs the diagnostic test of Walker et al. \cite{walker2013}. This diagnostic assesses the overlapping support of the distribution of the preference score, which is a transformation of the propensity score\footnote{Define the preference score $Pref$ as a transformation of the propensity score $p(t\g\mbx)$ that adjusts for the probability of treatment $p(t=1)$, 
 \begin{align*}
     \ln [Pref/(1-Pref] = \ln [p(t \g \mbx)/(1 - p(t \g \mbx)] - \ln [p(t=1)/ (1-p(t=1)].
 \end{align*}}, on the treatment and control groups.  If there is overlapping support (at least half the mass with preference between 0.3 and 0.7) then the study is said to be in \textit{equipoise}. If a study fails the diagnostic, then the analyst considers if an instrument has been missed and removes it, or interprets the results with caution.

\textbf{4. Match or stratify the dataset based on propensity scores and then check covariate balance.} Matching \citep{rubin1973, rubin1973a, stuart2010}  or subclassification \citep{rosenbaum1984} on propensity scores can be used to create groups of individuals who are similar. The details about the two methods are provided in Supplementary S1. The remaining of the algorithm is explained in the context of 1-to-1 matching. Once the matched groups are created, balance is assessed by computing the standardized mean difference (SMD) of each covariate between the treated group and the control group from the matched dataset
\begin{align*}
    \textrm{SMD} &= \frac{\bar{x}_{t=1} - \bar{x}_{t=0}}{\sqrt{(\sigma_{t=1}^2 + \sigma_{t=0}^2)/2}},
\end{align*}
where $\bar{x}_{t=1}$ and $\bar{x}_{t=0}$ are the mean of the covariate in the treated and the control group respectively, and $\sigma_{t=1}^2$ and $\sigma_{t=0}^2$ are the variance of the covariate in the treated and the control group respectively. Following Austin\cite{austin2009}, if any covariate has a SMD over 0.1 \citep{austin2009}, then the comparison is said to be out of balance, and the study needs to be discarded (or interpreted with caution).

\textbf{5. Estimate the causal effect.} The last step is to use the matched data to estimate the causal effect. In the simulations in Section \ref{sec:sim}, the causal effect of interest is the average treatment effect
\begin{align*}
    \textrm{ATE} = \E{Y_i(1)-Y_i(0)},
\end{align*} 
where $Y_i(1)$ and $Y_i(0)$ are the potential outcomes for a subject under treatment and under control. 

To estimate the ATE using matched data, a linear regression is fitted on the matched data with a treatment indicator variable. The coefficient of the treatment indicator is the average treatment effect. When subclassification is used to create balanced subclasses, the effect is estimated within each subclass and then aggregated across subclasses. The weight for each subclass is proportional to the total number of individuals in each subclass. 

In the empirical studies of Section \ref{sec:emp}, the causal effect of interest is hazard ratio (the outcome $Y$ is time to event). When matching is used, we fit a Cox proportional hazards model \citep{cox1972} on the matched dataset to estimate the hazards ratio. When subclassification is used, we fit a Cox model within each subclass and then weigh the conditional hazards ratio by the size of the subclass to obtain the marginal hazards ratio. More details are provided in Supplementary S2.


\subsection{Adjusting for indirectly measured confounders} \label{subsec:pinpoint}
As noted in the introduction, LSPS has been found to adjust for known but indirectly measured confounders \citep{hripcsak2020,schuemie2020a, chen2020}. We describe here an assumption under which LSPS will adjust for indirectly measured confounding.

\begin{figure}
    \centering
    \begin{subfigure}[b]{0.3\textwidth}
        \includegraphics[width=\linewidth]{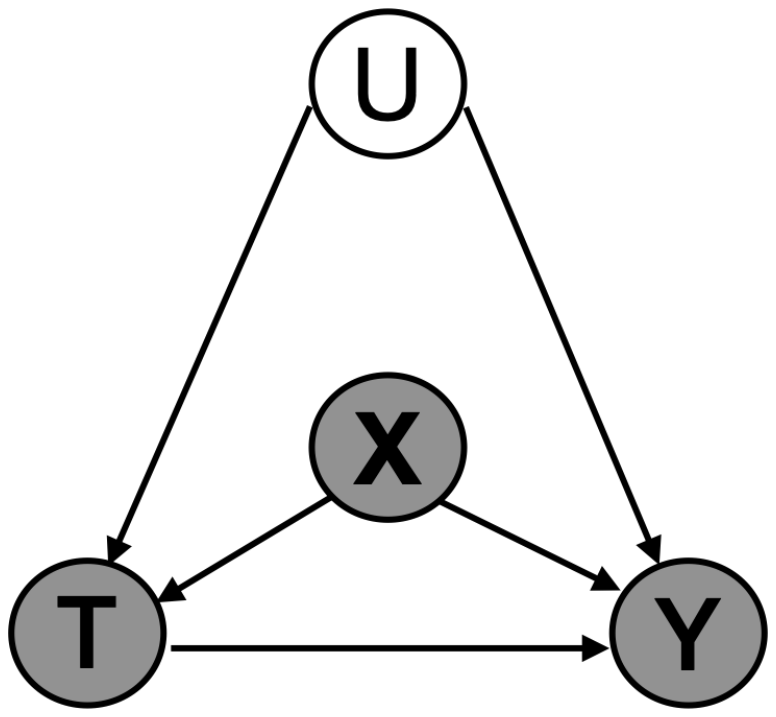}
        \caption{No pinpointability}
    \end{subfigure}
    \hfill
    \begin{subfigure}[b]{0.3\textwidth}
        \includegraphics[width=\linewidth]{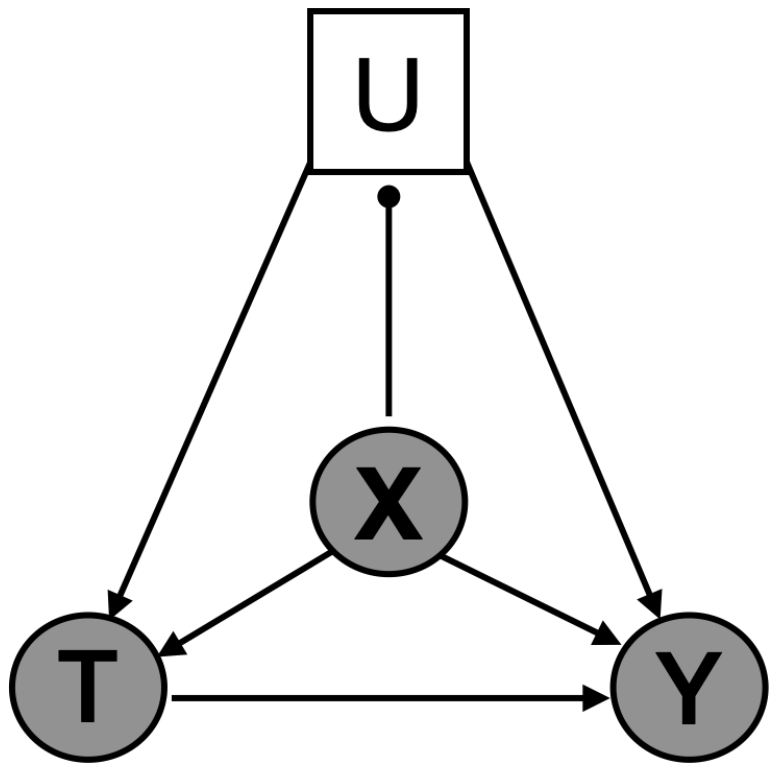}
        \caption{Perfect pinpointability}
    \end{subfigure}
    \hfill    
    \begin{subfigure}[b]{0.3\textwidth}
        \includegraphics[width=\linewidth]{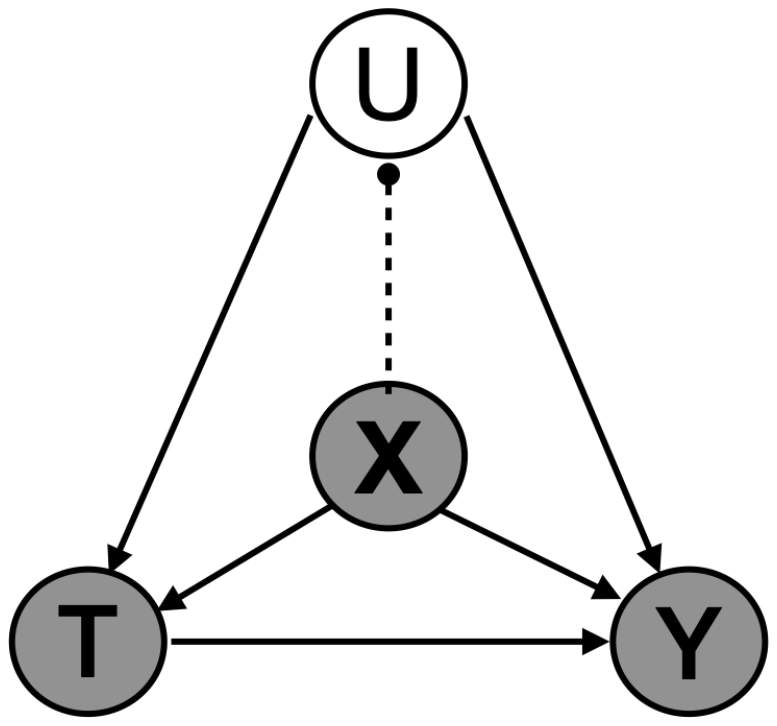}
        \caption{Weak pinpointability}
    \end{subfigure}
    \caption{Causal graphs to estimate the treatment effect of $T$ on the outcome $Y$. (a) Causal graph under no pinpointability. The unmeasured confounder $U$ is not pinpointed by $\mbX$. (b) Causal graph under perfect pinpointability. The unmeasured confounder $U$ is a deterministic function of the measured covariates $\mbX$. (c) Causal graph under weak pinpointability. The unmeasured confounder is only partially pinpointed by $\mbX$. Random variables are represented with circles, deterministic variables are represented with squares, measured variables are shaded, indirectly measured or unmeasured are not shaded, strong pinpointing is presented with a solid line, and weak pinpointing is presented with a dash line.}
\label{fig:gm}
\end{figure}

Consider the causal graph in Fig.\ref{fig:gm} for an individual $i$ (the subscript is omitted in the graph), where $T_i \in \{0,1\}$ is a binary treatment, $Y_i$ is the outcome (either binary or continuous), $\mbX_i \in \{0,1\}^M$ is a high-dimensional vector of observed pre-treatment covariates with length M (which includes observed confounders and other variables), and $U \in \R$ is the indirectly measured confounder. The goal is to estimate the causal effect of treatment $T$ on the outcome $Y$. To do so, we need to adjust for both pre-treatment covariates $\mbX$ (including directly measured confounders) and $U$ (indirectly measured confounder). We assume that there are no other unmeasured confounders.

In the following sections, we will demonstrate that LSPS can still produce unbiased causal estimates even in the presence of indirectly measured confounders. We first introduce Assumption 1, which indicates the relationship between measured covariates and indirectly measured confounders. 

\begin{assumption} (Pinpointability of indirectly measured confounder) An unmeasured confounder $U$ is said to be pinpointed by measured covariates if
\begin{align}
  \rmp(u \g \mbx) = \delta(f(\mbx))
\end{align}
where $\delta(\cdot)$ denotes a point mass at $f(\cdot)$.
\end{assumption}

In other words, the indirectly measured confounder $U$ can be represented by a deterministic function $f$ of the measured covariates $\mbX$. Theorem 1, building upon Assumption 1, formally states the conditions that LSPS needs in order to obtain unbiased causal estimates by only conditioning on measured covariates. We use the potential outcome framework by Rubin \cite{rubin1974}. Let $Y_i(1)$ and $Y_i(0)$ denote the potential outcome under treatment and under control respectively for an individual $i$.

\begin{theorem} The treatment and the potential outcomes are independent conditioning on all confounders, both the directly measured ($\mbX$) and indirectly measured ($U$),
\begin{align}{}
T \indpt Y(1), Y(0) \g \mbX, U.
\end{align}
Under the pinpointability assumption, the above conditional independence can be reduced to only conditioning on the measured covariates,
\begin{align}{}
T \indpt Y(1), Y(0) \g \mbX.
\end{align}
\end{theorem}

In other words, the causal effect of the treatment on the outcome is identifiable by only adjusting for the measured covariates $\mbX$. We do not need to know the indirectly measured confounders $U$ or its functional form $f(\cdot)$.

\begin{proof} Theorem 1 relies on the marginalization over $U$ in computing the propensity score using high-dimensional measured covariates,  
\begin{align}
  \rmp(t \g \mbx) &= \int \rmp(t \g u, \mbx)  \rmp(u \g \mbx)  \dd u \\
  &= \rmp(t \g u^*, \mbx),
\end{align}
where $u^* = f(\mbx)$.
\end{proof}

When $U$ is weakly pinpointed by $\mbX$, or in other words, $f(\mbX)$ measures $U$ with error, the average treatment effect is not point identifiable. Assuming identification holds conditional on the unmeasured confounder, Ogburn and VanderWeele \cite{ogburn2013} show that the average treatment effect adjusting for the noisy measured confounder is between the unadjusted and the true effects, under some monotonicity assumptions. We extend the work by Ogburn and VanderWeele \cite{ogburn2013} to conditions where additional confounders and covariates exist. 

\begin{theorem} Let $T$ be a binary treatment, $Y$ be an outcome, $\mbX_C$ be all measured confounders, $U$ be an ordinal unmeasured confounder, and $U'$ be the noisy measurement of $U$. Assume that the measurement error of $U$ is tapered and nondifferential with respect to $T$ and $Y$ conditional on $\mbX_C$, and that $\E{Y \g T, U, \mbX_C}$ and $\E{T \g U, \mbX_C}$ are monotonic in $U$ for each value $\mbx_C$ in the support of $\mbX_C$. Then the average treatment effect adjusting for the measured covariates lies between the true effect and effect adjusting for only the measured confounders, that is, $ATE_{true} \leq ATE_{cov} \leq ATE_{conf}$ or $ATE_{true} \geq ATE_{cov} \geq ATE_{conf}$, where  $ATE_{true} = \EE{X_C, U}{Y(1)} - \EE{X_C,U}{Y(0)}$, $ATE_{cov} = \EE{X_C, U'}{Y \g T=1} - \EE{X_C, U'}{Y \g T=0}$, and $ATE_{conf} = \EE{X_C}{Y \g T=1} - \EE{X_C}{Y \g T=0}$.
\end{theorem}

Theorem 2 states that the average treatment effect adjusting for all measured covariates (i.e., by LSPS) is bounded between the true effect and the effect adjusting for only the measured confounders. Nondifferential misclassification error means that $p(U=u’ | U=u, T=t, Y=y, \mbX_C=\mbx_c) = p(U=u’ | U=u, T=t’, Y=y’, \mbX_C=\mbx_c’)$ for all $y, y' \in \mathcal{Y}$, $t, t' \in \{0,1\}$, and $\mbx_c, \mbx_c' \in \mathcal{X_C}$. Tapered misclassification error means that the misclassification probabilities $p_{ij} = p(U' = i \g U = j)$, $i, j \in \{1, \dots, K\}$ is nonincreasing in both directions away from the true unmeasured confounder.

We show that Theorem 1 and Theorem 2 hold in the simulations with various degrees of pinpointability. Proof of Theorem 2 is an extension of the work by Ogburn and VanderWeele \cite{ogburn2013}, and is given in the Supplement.

\subsection{Effect of pinpointing on instruments and M-bias}
Because LSPS uses a large number of covariates, there is a concern that adjusting for these covariates will induce bias due to M-structure colliders, instrumental variables (IVs), and near instrumental variables (near-IVs). As noted above, our goal is not to do instrumental variable analysis but rather to remove their potential effect of increasing variance and amplifying bias. IVs are addressed in part by domain knowledge and diagnostics, but some IVs may remain. In this section, we discuss how LSPS in the setting of pinpointing may address them.

\begin{figure}
     \centering
     \begin{subfigure}[b]{0.3\textwidth}
         \centering
         \includegraphics[width=\textwidth]{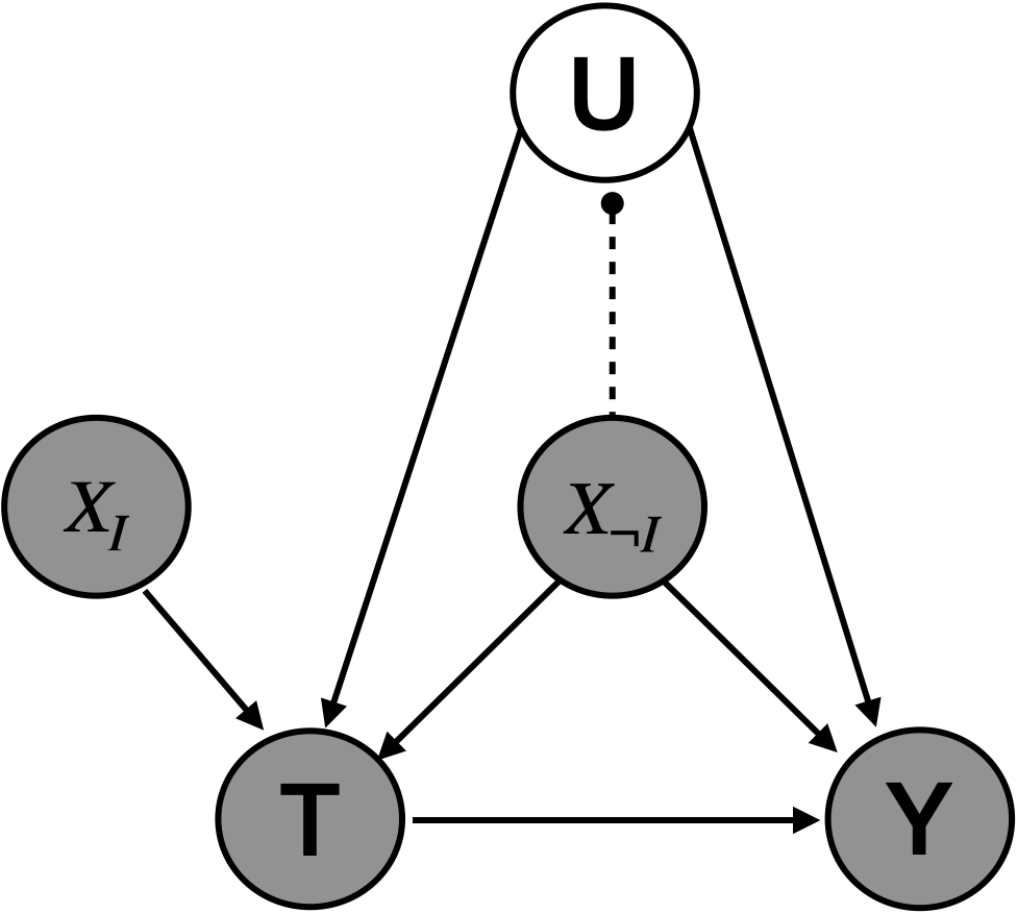}
         \caption{IV}
         \label{fig:iv}
     \end{subfigure}
     \hfill
     \begin{subfigure}[b]{0.3\textwidth}
         \includegraphics[width=\textwidth]{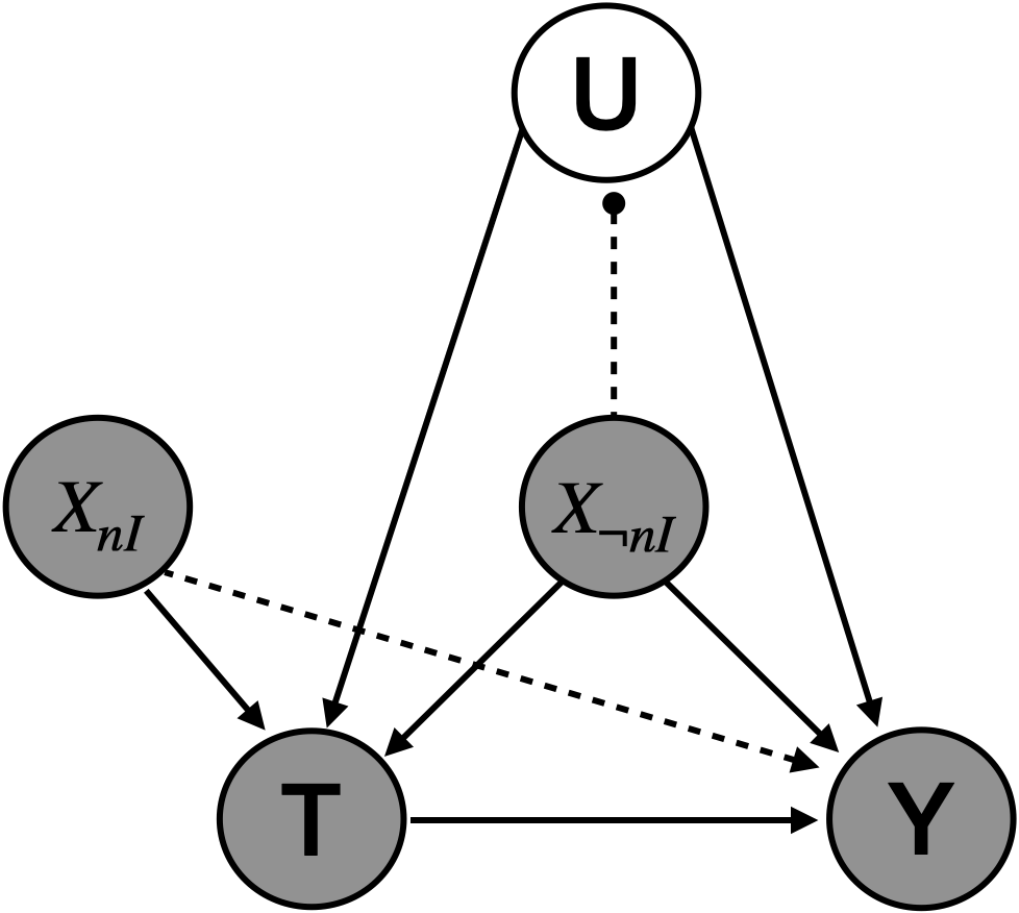}
         \caption{near-IV}
         \label{fig:neariv}
     \end{subfigure}
     \hfill
     \begin{subfigure}[b]{0.3\textwidth}
         \centering
         \includegraphics[width=0.77\textwidth]{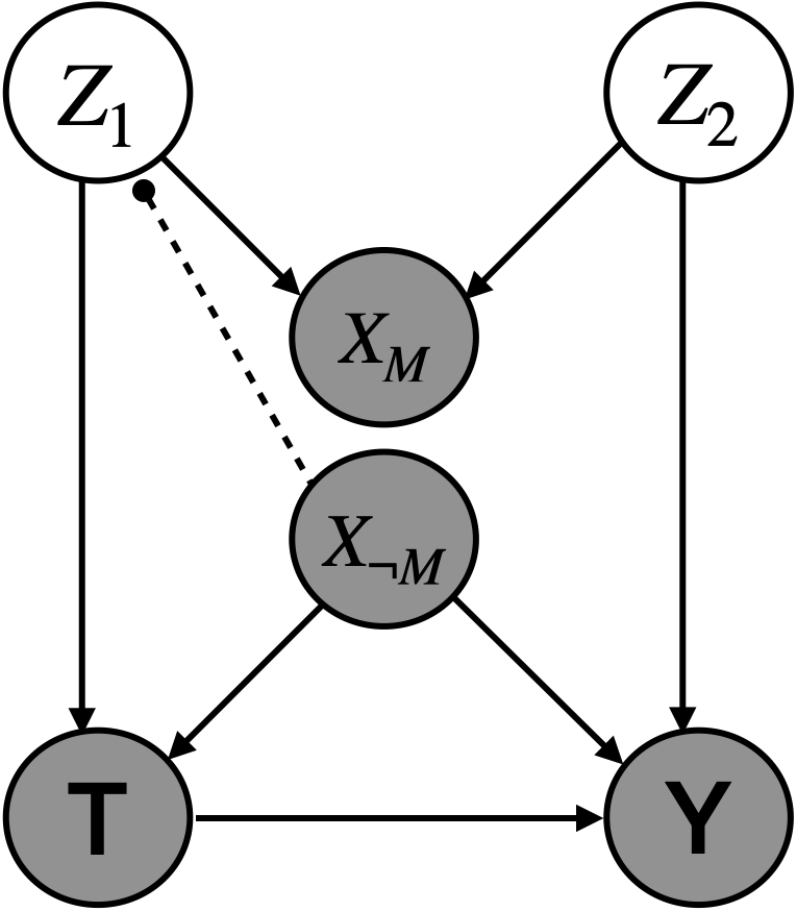}
         \caption{M-structure collider}
         \label{fig:mbias}
     \end{subfigure}
    \caption{Causal graph of a) instrumental variable, b) near-instrumental variable and c) M-structure collider. The dashed line with a solid dot means that the variable by the solid dot can be pinpointed by the measured covariates. We use different subscripts to distinguish the measured covariates playing different roles in the causal graph.}
    \label{fig:properties}
\end{figure}

\subsubsection{Effect on IV and near-IV} \label{sec:IV}
Instrumental variables \citep{hernan2006} may persist despite LSPS's procedures. In the setting of unmeasured confounding, IV can cause bias amplification as shown numerically \citep{bhattacharya2007, middleton2016} and proved theoretically in various scenarios \citep{pearl2010, pearl2011, pearl2013, wooldridge2016, steiner2016, ding2017}. Insofar as pinpointing adjusts for indirectly measured confounding (Fig.\ref{fig:iv}), even if there are IVs in the propensity score model, they will not produce bias amplification \citep{steiner2016}.

Near-instrumental variables (near-IVs) \citep{myers2011}, which are weakly related to the outcome and strongly related to the treatment, may also lead to bias amplification \citep{myers2011, pearl2010, steiner2016}, and the bias amplification or the confounding may dominate. Just as for IVs, pinpointing (Fig.\ref{fig:neariv}) may reduce bias amplification by reducing indirectly measured confounding \citep{steiner2016}, while the confounding is eliminated by adjusting for the near-IV.

\subsubsection{Addressing M-bias}
Despite the use of pre-treatment variables, bias through colliders is still possible due to causal structures like the one in Fig. \ref{fig:mbias}, known as an M-structure, causing M-bias. In this case, two unobserved underlying causes create a path from $T$ to $Y$ via a collider that can precede $T$ in time. If the collider is included in the many covariates, then this can induce bias. LSPS may be able to address M-bias in the following way. If the common cause between the treatment and the collider ($Z_1$) can be pinpointed by the measured covariates, then this will block the back-door path from $T$ to $Y$. Similarly, the common cause between the outcome and the collider ($Z_2$) could be pinpointed, also blocking the path. The assertion that one or both of these common causes is pinpointed is similar to the assertion that $U$ is pinpointed.



\section{Simulations}\label{sec:sim}
We use the simulation to show that, under the assumption of pinpointability, LSPS can adjust for the indirectly measured confounding, and as the condition deviates from pinpointability, bias in LSPS increases, and the estimate by LSPS is between the true effect and the effect adjusting for only measured confounders. In this simulation, we assume that the large number of covariates $\mbX$ are derived from a smaller number of underlying latent variables $\mbV$. This data generating process induces dependencies among the measured covariates, mimicking the dependencies observed in EHR.


\begin{figure}
    \centering
    \includegraphics[width=0.3\textwidth]{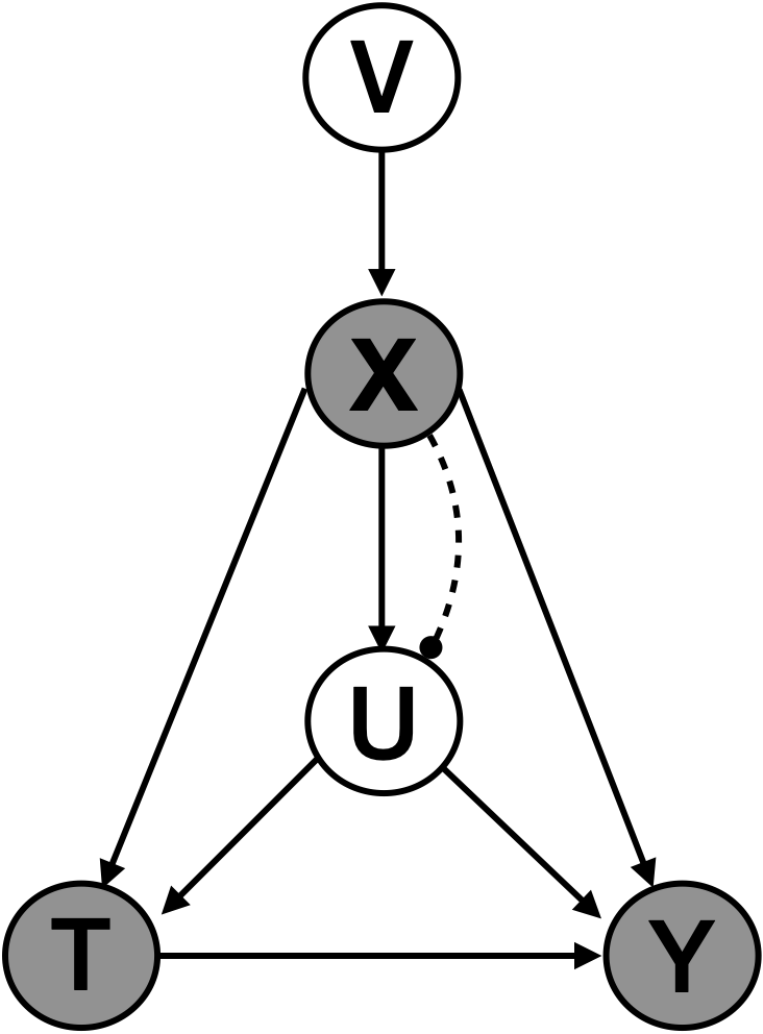}
    \caption{Causal diagram of the simulation to estimate the effect of the treatment $T$ on the outcome $Y$. The high-dimensional measured covariates $\mbX$ are induced by a low-dimensional latent variable $\mbV$. The unmeasured confounder $U$ is simulated as a function of the measured covariates $\mbX$. When the function is deterministic, $U$ is pinpointed by $\mbX$. When the function is stochastic, $U$ is weakly pinpointed by $\mbX$. The degree of pinpointability is varied by adding varied amount of noise into the function.}
\end{figure}

\subsection{Simulation setup}
Each simulated data set contains $N=5,000$ patients, $M=100$ measured covariates (including $10$ measured confounder), $1$ indirectly measured confounder, $10$ latent variables, a treatment and an outcome. The data set is $\cD = \{\mbv_i, \mbx_i, u_i, t_i, y_i\}_{i=1}^N$, where $\mbv_i$ is a vector of latent variables, $\mbv_i = (v_{i1}, \dots, v_{i10})$; $\mbx_i$ is a vector of measured covariates, $\mbx_i = (x_{i1}, \dots, x_{iM})$; $u_i$, $t_i$ and $y_i$ are all scalar, representing the indirectly measured confounder, treatment and outcome respectively.

Fig.3 shows the data generating process using a causal diagram. Below are the steps to simulate data for patient $i$.

\begin{enumerate}
    \item Simulate the latent variable $\mbv_i$ as 
    $$\mbv_i \sim \text{Bernoulli}(0.5)^K.$$
    \item Simulate measured covariates $\mbx_i$ as
    $$\mbx_i \sim \text{Bernoulli}(\text{sigmoid}(\mbv_i^\top\mb\beta_x)),$$
    where $\mb\beta_x \sim \cN(0,0.1)^{K \times M}$.
    \item Simulate the indirectly measured confounder $u_i$ as 
    $$u_i = \mbx_i^\top\mb\beta_u,$$
    where $\mb\beta_u \sim \cN(0,1)^M$. Notice that $u$ is a deterministic function of $\mbx$. To allow only a small subset of the covariates pinpoint $u$, we randomly select 90\% of the $\mb\beta_u$ and set their value to 0.
    \item Simulate the treatment $t_i$ as
    $$t_i \sim \text{Bernoulli}(\text{sigmoid}(\mbx_i^\top\mb\gamma_x  + u_i\gamma_u)),$$
    where the effect of the indirectly measured confounder on the treatment $\gamma_u = 1$, $\mb\gamma_x \sim \cN(0.5, 1)$ for the 10\% of covariates that serve as measured confounders. The rest of the $\gamma_x$ are set to 0. 
    \item Simulate the outcome $y_i$ as 
    $$y_i \sim \cN(\mbx_i^\top \mb\eta_x + u_i\eta_u + t_i\nu, \ 0.1),$$
    where the true causal effect $\nu = 2$, the effect of the indirectly measured confounder on the outcome $\eta_u = 1$, $\mb\eta_x \sim \cN(0.5, 1)$ for the covariates that serve as measured confounders and 0 otherwise. 
\end{enumerate}

The above steps illustrate the simulation under pinpointability. To increase the deviation from pinpointability, we add an increasing amount of random noise to the indirectly measured confounder. To do so, we modify the simulation of $u_i$ in Step 3 to be
\begin{align*}
    u_i = \mbx_i^\top\mb\beta_u + \epsilon_i
\end{align*}
 where $\epsilon_i \sim \cN(0, \sigma^2)$. To increase deviation from pinpointability, we increase $\sigma^2$ from $10^{-4}$ to $10^4$. At each pinpointability level, the Gaussian noise is non-differential with respect to the treatment, outcome, and the measured covariates, and tapered (decreasing in both directions away from the true unmeasured confounder). We simulate 50 datasets at each pinpoitability level.

\subsection{Statistical analysis} \label{subsec:eval}
We demonstrate LSPS’s capacity in adjusting for unmeasured confounding relative to other methods under varied degree of pinpointability. Specifically, we compared the following five methods:
\begin{itemize}
    \item unadjusted: no covariate was adjusted for.
    \item manual without U: adjust for all confounders not including $U$
    \item manual (oracle): adjust for all confounders including $U$
    \item LSPS without U: adjust for all measured covariates not including $U$
    \item LSPS: adjust for all covariates including $U$
\end{itemize}

Notice that manual with $U$ coincides with oracle in this simulation, because manual with $U$ adjusts for nothing but the confounders, both measured and unmeasured. In practice, because the confounding structure is rarely known, it is unlikely a manual method captures all the confounders.

For the four methods that adjust for confounders, we estimated propensity scores with L1-regularized logistic regression (and selected the regularization parameter with cross-validation). We then used 1:1 matching and subclassification (results not shown) to create a balanced dataset. To estimate the average treatment effect, we fit a linear regression model on the balanced dataset. We then calculated the mean, 95\% confidence interval, and root-mean squared error (RMSE) of the effect estimates. 

The RMSE, defined as follows, can be calculated because the true treatment effect is known in the simulation ($\nu = 2$). This metric is not applicable to the empirical study because the true effect of medications is unknown in practice. 
\begin{align*}
    \text{RMSE} &= \sqrt{\frac{1}{S}\sum_s (\hat{\nu}_s - \nu)^2},
\end{align*}
where $\hat{\nu}_s$ is the effect estimate at simulation $s$, and we simulate a total of $S=50$ datasets at each given pinpointability condition.

\subsection{Results}
\begin{figure}
    \centering
    \includegraphics[width=0.9\textwidth]{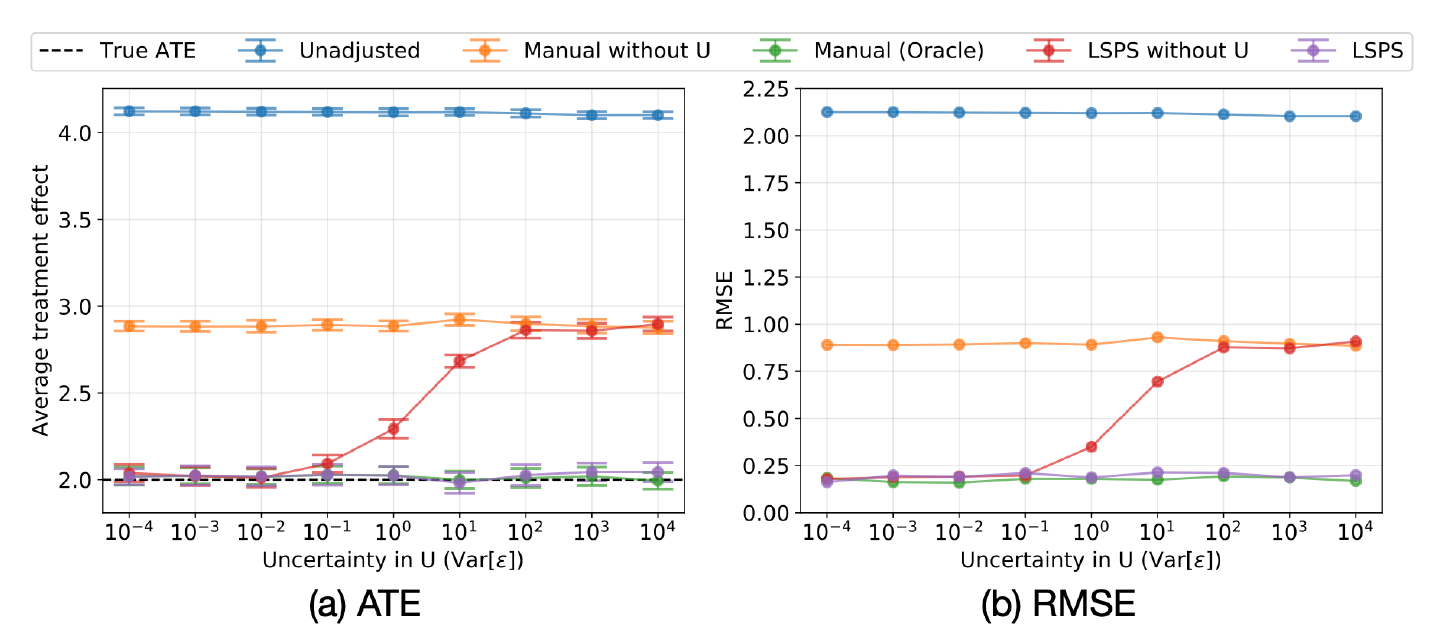}
    \caption{Sensitivity analysis of pinpointability in Simulations 1. As pinpointability of the indirectly measured confounder decreases, LSPS's ability to adjust for the indirectly measured confounder decreases. (a) The mean and 95\% CI of the estimated average treatment effect. (b) The RMSE of the estimated average treatment effect. 
    } 
    \label{fig:sim1res}
\end{figure}

Fig.~\ref{fig:sim1res} shows the results of the simulation. When pinpointability holds reasonably well, the unmeasured confounder has a bigger impact on the manual method than on LSPS. As pinpointability gets weaker, adjusting for large-set of covariates becomes less adequate for accounting for the unmeasured confounder, and the estimated treatment effect approaches the estimate from the method adjusting for only the measured confounders.

Under strong pinpointability, the estimates from the two large-scale approaches (LSPS without $U$ and LSPS) have almost the same bias, variance, and RMSE compared to the estimate from the oracle. The manual approach (manual without $U$) does not benefit from pinpointability because it does not include covariates that assist pinpointing. As the condition deviates from strong pinpointability, the large-scale approach (LSPS without $U$) becomes increasingly biased, approaches and eventually overlaps the estimate from the manual without $U$ approach. This simulation result matches Theoreom 2 that $ATE_{cov}$ is between $ATE_{conf}$ and $ATE_{true}$ under certain monotoniticty assumptions. In the simulation, $ATE_{conf}$ is given by manual without $U$, $ATE_{cov}$ is given by LSPS without $U$, and $ATE_{true}$ is given by manual with $U$.

\section{Empirical studies}\label{sec:emp}

We now use real data to compare LSPS to the traditional propensity-score adjustment (with manually selected covariates) to adjusting for confounding. With an EHR database, we compared the effect of two anti-hypertension drugs, hydrochlorothiazide and lisinopril, on two clinical outcomes, acute myocardial infarction (AMI) and chronic kidney disease (CKD). For both outcomes, type 2 diabetes mellitus (T2DM) is a known confounder. Thus, by including or excluding T2DM in an adjustment model while keeping other covariates the same, we can assess a method's capacity in adjusting for a known confounder that is not directly measured but may be correlated with measured covariates. 

\subsection{Cohort and covariates}\label{sec:cohort}
We used a retrospective, observational, comparative cohort design \cite{hernan2016}. We included all new users of hydrochlorothiazide monotherapy or lisinopril monotherapy and defined the index date as the first observed exposure to either medication. We excluded patients who had less than 365 days of observation prior, a prior hypertension treatment, initiated another hypertension treatment within 7 days, or had the outcome prior to index date. We followed patients until their end of continuous exposure, allowing for maximum gaps of 30-days, or their end of observation in the database, whichever came first. 

For the LSPS-based approach, we used more than 60,000 covariates in the EHR database, including demographics, all medications in the 365 days prior to index date, all diagnoses in the 365 days prior to index date, and the Charlson Comorbidity Index score, as baseline covariates in the propensity model. 

For traditional PS adjustment, covariates were selected by experts for inclusion in related hypertension drug studies \citep{chien2015,hicks2018, ku2018, magid2010, hasvold2014}, including
T2DM, anti-glycemic agent, age groups, female, index year,  coronary artery disease, myocardial infarction, asthma, heart failure, chronic kidney disease, atrial fibrillation, Charlson index - Romano adaptation, Platelet aggregation inhibitors excl. heparin, Warfarin, corticosteroids for systemic use, dipyridamole, non-steroidal anti-inflammatory drugs(NSAIDS), proton-pump inhibitors (PPIs), statins, estrogens, progestogens, body mass index (BMI), chronic obstructive pulmonary disease(COPD), liver disease, dyslipidemia, valvular heart disease, drug abuse, cancer, HIV infection, smoking and stroke.

Most covariates (e.g., diagnoses, medications, procedures) were encoded as binary, that is, 1 indicates the code is present in the patient’s medical history prior to treatment, and 0 otherwise. For some variables that are often considered as continuous (e.g., lab tests), LSPS does not impute the value because imputation could do more harm than good when missing mechanism is not known. Instead, LSPS encodes the lab ordering pattern as binary variables. Residual error appears as measurement error.

The study was run on the Optum\textcopyright{} de-identified electronic health record database of aggregated electronic health records.

\subsection{Statistical analysis}

We examined a method's capacity in adjusting for confounding that is not directly measured by comparing the effect estimates from each method with or without access to the confounder. We excluded all variables related to T2DM, including diagnoses and anti-glycemic medications in models without access to indirectly measured confounders. We included an unadjusted method as a baseline for comparison. Specifically, we studied the following five methods (analogous to the five methods in the simulation):
\begin{itemize}
    \item unadjusted: no covariate was adjusted for.
    \item manual: adjust for a list of manually selected confounders.
    \item manual without T2DM: adjust for a list of manually selected confounders without T2DM-related confounders. 
    \item LSPS: adjust for all pre-treatment covariates in the database.
    \item LSPS without T2DM: adjust for all pre-treatment covariates in the database without T2DM-related confounders. 
\end{itemize}
For the four methods that adjust for confounders, we estimated propensity scores with L1-regularized logistic regression (and selected the regularization parameter with cross-validation). We used subclassification and stratified the dataset into 10 subclasses. To estimate the treatment effect, we fit a Cox proportional-hazards model \citep{cox1972} to estimate the hazard ratio (HR). We then calculated the mean and 95\% confidence interval of the HR.

\subsection{Results}
\begin{figure}
     \centering
     \begin{subfigure}[b]{0.45\textwidth}
         \includegraphics[width=\textwidth]{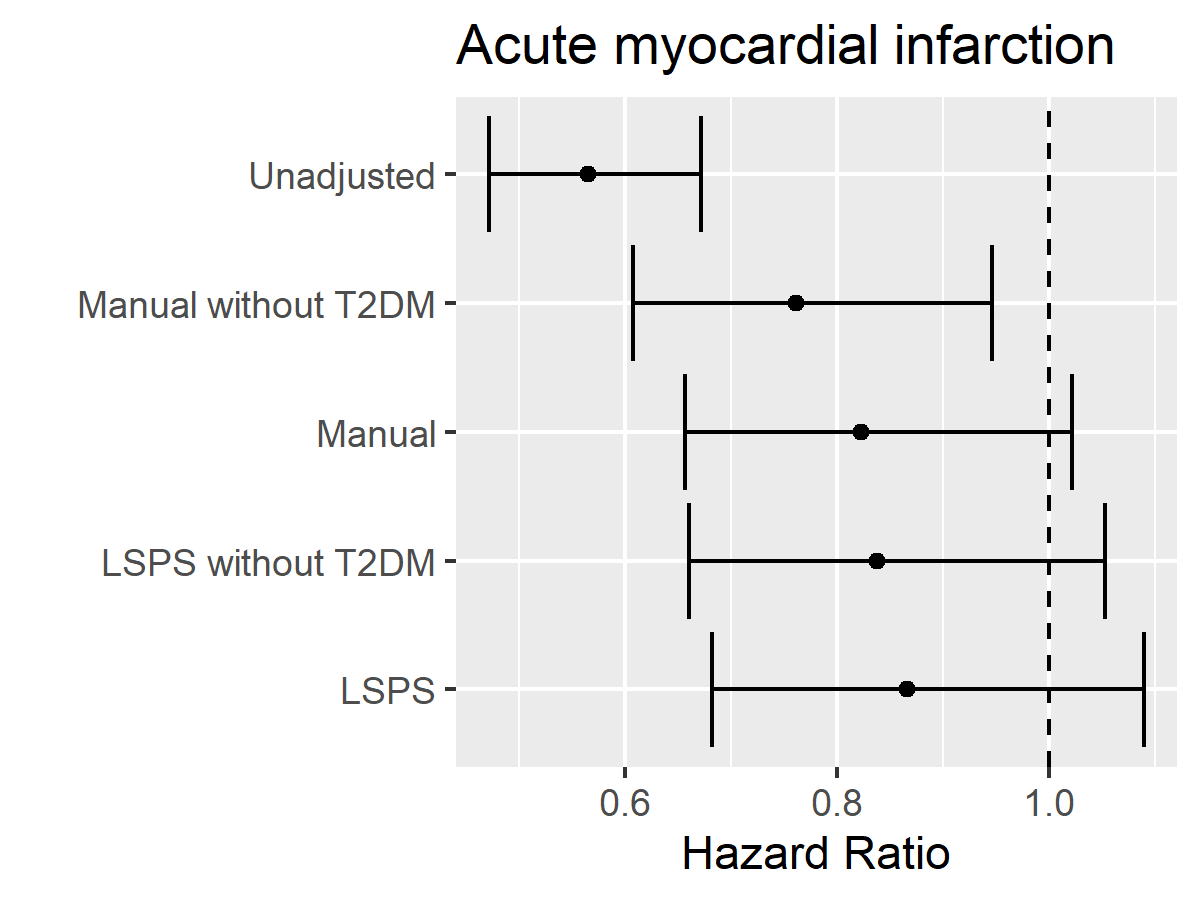}
         \caption{Acute myocardial infarction}
         \label{fig:HRAMI}
     \end{subfigure}
     \hfill
     \begin{subfigure}[b]{0.45\textwidth}
         \includegraphics[width=\textwidth]{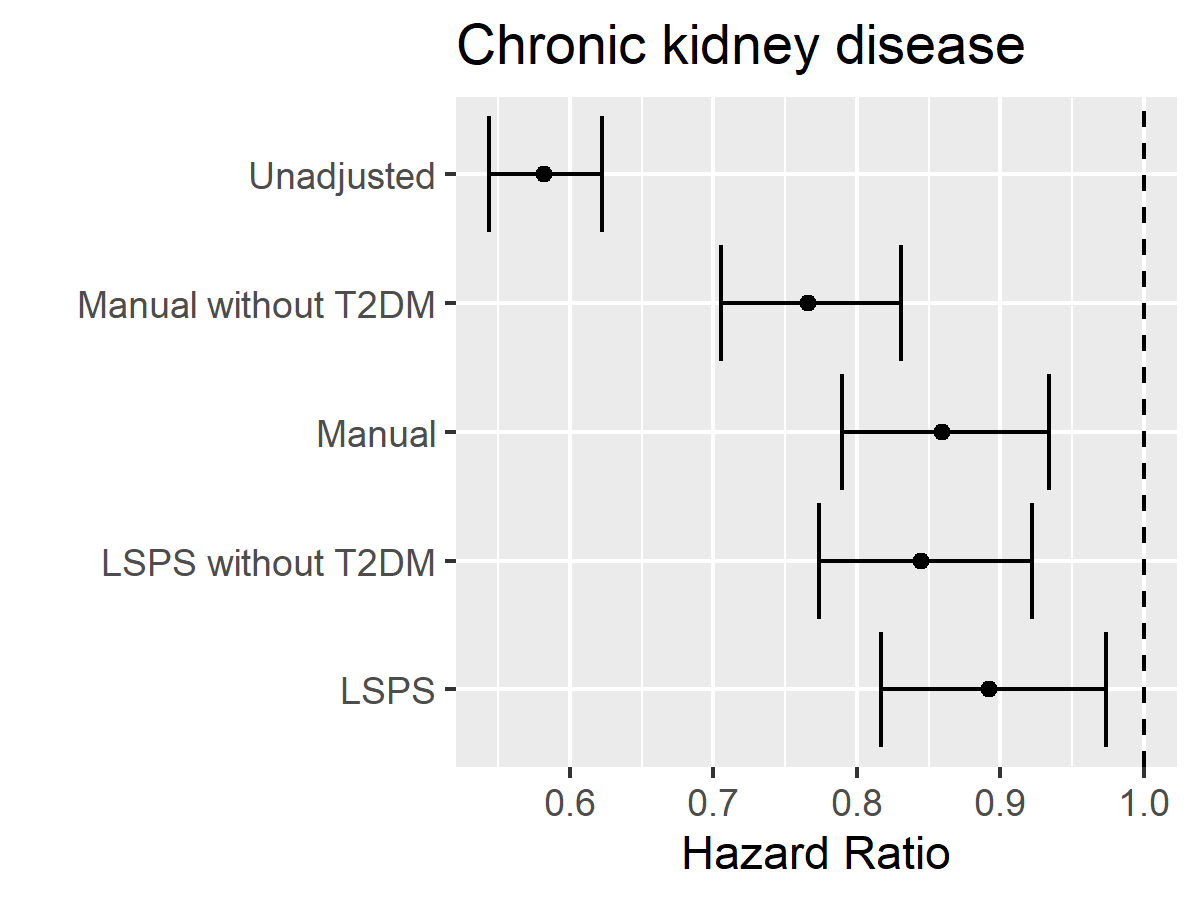}
         \caption{Chronic kidney disease}
         \label{fig:HRCKD}
     \end{subfigure}
\caption{Comparison of hazard ratio from the unadjusted model and four models adjusting for confounders. The indirectly measured (or unused) confounder T2DM had a bigger impact on the HR estimated by manual models than by LSPS. (a) HR of the two anti-hypertensive medications on AMI. (b) HR of the two anti-hypertensive medications on CKD.}
\label{fig:hr}
\end{figure}

Fig.~\ref{fig:hr} shows the results of empirical studies. These results show that the T2DM had a bigger impact on the manual methods than on the LSPS methods. The impact was determined by comparing the absolute difference in effect estimates between the two manual models versus the two LSPS-based models. In the CDK study, the absolute difference between the two manual methods was 0.09 (Manual without T2DM: HR 0.77 [95\% CI, 0.71-0.83]; Manual: HR 0.86 [95\% CI, 0.79-0.93]), higher than the absolute difference between the two LSPS-based methods, which was 0.05 (LSPS without T2DM:  HR 0.84 [95\% CI, 0.77-0.92]; LSPS: HR 0.89 [95\% CI, 0.82-0.97]). In fact, the estimates for manual with T2DM, LSPS with T2DM, and LSPS without T2DM were all closer to each other than to manual without T2DM. Therefore, whether manual with T2DM or LSPS with T2DM is actually closer to ground truth, LSPS without T2DM is closer to either one than is manual without T2DM.

This finding suggests that by including large-scale covariates, one has a better chance of correcting for confounders that are not directly measured. The pinpointability assumption is more likely to hold when there are many measured covariates. 

\section{Discussion}
We have illustrated conditions under which LSPS adjusts for indirectly measured confounding and the impact of violations of such conditions on effect estimation. We have found in previous practice, in our current simulations, and in our current real-world study that indirectly measured (or unused) confounding can be adjusted for in LSPS, apparently working better than smaller, manually engineered sets of covariates that are also missing the confounder.

Even though pinpointing in the current simulation is achieved by generating the unmeasured confounder from a function of the measured covariates, this does not suggest the causal direction between the unmeasured confounder and the measured covariates. In medicine, it is likely that an unmeasured confounder (e.g., a disease such as T2DM) induces dependencies among large-scale clinical covariates (e.g., medications for treating the disease, laboratory tests for monitoring the disease, and other diseases that often co-occur with the disease can be correlated). In other words, the unmeasured confounder could be a latent variable in a factor model, and the strength of pinpointing depends on the number of measured covariates and the degree of dependency among the covariates.

We describe here methods that are related to LSPS. These methods are related to LSPS in different ways. Section \ref{subsec:proxyetc} compares and contrasts LSPS to other methods that also address unmeasured confounding in causal effect estimation. Section \ref{subsec:hdps} compares LSPS to another propensity score-based method that also uses large-scale covariates. Section \ref{subsec:deconfounder} draws similarity between LSPS and another method for causal effect estimation in the presence of indirectly measured confounding where pinpointability is a required assumption. 

\subsection{Relation to proxy variable, multiple imputation and residual bias detection}\label{subsec:proxyetc}
Studies such as those by Kuroki and Pearl \cite{Kuroki2014Measurement}, Miao et al.\cite{miao2018}, and Tchetgen Tchetgen et al.\cite{Tchetgentchetgen2020} have shown that causal effects can be identified by observing proxy variables of confounders that are not directly measured. In this case, the confounder is known but not measured, there is sufficient knowledge of the structural causal model such that proxies can be selected, and there is the knowledge that there are no other unmeasured confounders. In contrast, LSPS does not require explicit knowledge of the causal model.

Another approach is to use measured covariates to explicitly model unmeasured confounders using multiple imputation \citep{albogami2020, albogami2021}. While it differs from our approach, it exploits the same phenomenon, that some covariates contain information about unmeasured confounders. This is in contrast to LSPS, where there is no explicit model of unmeasured confounders; adjusting for measured covariates should be effective for causal inference as long as the measured covariates pinpoint the unmeasured confounders.

Given the need to assume no additional unmeasured confounding—additional in the sense of not being pinpointed or not having proxies—a complementary approach is to estimate the degree of residual bias, potentially including additional unmeasured confounding. Large-scale use of negative and synthetic positive controls \citep{schuemie2018a,schuemie2018} can detect residual bias and can additionally be used to calibrate estimates to assure appropriate coverage of confidence intervals. LSPS is usually coupled with such empirical calibration \citep{schuemie2020b, schuemie2018a}. 

\subsection{Relation to high-dimensional propensity score adjustment} \label{subsec:hdps}
LSPS adjusts for all available pre-treatment covariates. In practice, because the sample size is limited, regularized regression selects a subset of variables to represent the information contained in the whole set of covariates, but the goal is to represent all the information nonetheless. Therefore, LSPS diagnostics \citep{schuemie2018a, tian2018} test balance not just on the variables that regularized regression included in the model, but on all the covariates. All covariates are retained because even those that are not direct confounders may still contribute to the pinpointing of the unobserved confounders. Therefore, LSPS is not a confounder selection technique.

LSPS is distinct from techniques that attempt to select confounders empirically \citep{schneeweiss2009}. Some of these techniques also start with large numbers of covariates, but they attempt to find the subset that are confounders using information about the treatment and outcome. They then adjust for the selected covariates. As long as all confounders are observed and then selected, adjusting for them should eliminate confounding. It may not, however, benefit from the pinpointing that we identify in this paper. Unlike LSPS, confounder selection techniques are dependent on the outcomes, and the outcome rates are often very low in medical studies, potentially leading to variability in selection. Empirical studies \citep{chen2020} show that adjusting for a small number of confounders does not successfully adjust for unobserved confounders, and an empirical comparison of the methods favored LSPS \citep{tian2018}.

\subsection{Relation to the deconfounder} \label{subsec:deconfounder}
LSPS and the deconfounder \citep{wang2019,wang2020} are distinct but share several features. The deconfounder is a causal inference algorithm that estimates unbiased effects of multiple causes in the presence of unmeasured confounding. Under the pinpointability assumption (unmeasured confounders are pinpointable by multiple causes), the deconfounder can infer unmeasured confounders by fitting a probabilistic low-rank model to capture the dependencies among multiple causes. The deconfounder has been applied to EHR data for treatment effect estimation in the presence of unmeasured confounding \citep{zhang2019a}. Both methods thus can be shown to address unmeasured confounders when there is pinpointing.

\section{Conclusions}
In summary, LSPS is a confounding adjustment approach that includes large-scale pre-treatment covariates in estimating propensity scores. It has previously been demonstrated that LSPS balances unused covariates and can adjust for indirectly measured confounding. This paper contributes to understanding conditions under which LSPS adjusts for indirectly measured confounders, and how causal effect estimation by LSPS is impacted when such conditions are violated. We demonstrated the performance of LSPS on both simulated and real medical data.

\section*{Acknowledgements}
This work was supported by NIH R01LM006910, c-01; ONR N00014-17-1-2131, N00014-15-1-2209; NSF CCF-1740833; DARPA SD2 FA8750-18-C-0130; Amazon; NVIDIA; and Simons Foundation.



 \bibliographystyle{elsarticle-num} 
 \bibliography{biblio}

\newpage
\section*{Supplementary Materials}
\section*{S1. Matching and Subclassification}

Both matching and subclassification can be used to create groups of similar individuals in Step 4 of LSPS.

\subsection*{S1.1 Matching}
There are a few criteria that need to be pre-specified for matching. First, the maximum allowed difference between matched individuals, often called caliper. A caliper can be defined at three commonly used scales: propensity score scale, standardized scale, and standardized logit scale. On the standardized scale, the caliper is defined as standard deviations of the propensity score distribution. The standardized logit scale is similar, except that the propensity score is transformed to the logit scale because the PS is more likely to be normally distributed on that scale (Austin, 2011). The default caliper is 0.2 on the standardized logit scale, as recommended by Austin (2011).

Second, the maximum number of persons in the control group to be matched to each person in the treatment group. In its simplest form, 1:1 nearest neighbor matching selects one individual from the control group with the smallest propensity score difference from a given individual in the treatment group. Alternatively, 1:k matching and varied ratio matching can be used. In this study, 1:1 matching is used.  

\subsection*{S1.2 Subclassification}
Alternative to matching, researchers can use the propensity model to stratify the dataset (Rosenbaum and Rubin, 1984). Subclassification divides individuals into subclasses within which the propensity scores are relatively similar. 

Researchers need to determine the number of subclasses and the boundaries of the subclasses. Current convention is to create 5-10 subclasses with equal support (or equal sample size), though more subclasses may be feasible and appropriate if sample size is large. We use 10 subclasses with equal support in this study. Each datapoint is then assigned to a subclass based on its propensity score. 

Once the subclasses are defined, LSPS checks that the subclassification achieves covariate balance. Balance is achieved if when we reweigh the data according to the subclasses, each covariate in the treatment and control group follows the same distribution.

To check covariate balance, we first compute the weight for each datapoint $w_i$ to be equal to the reciprocal of the number of datapoints from its treatment group within its subclass. Mathematically, that is,
\begin{align*}
    w_i &= 1/\Big(\sum_i^N 1(T_i=t)1(S_i=s)\Big),
\end{align*}
where the denominator is the number of datapoints that received treatment $t$ in subclass $s$. Then, the weighted mean of the covariate for treatment group $t$ is 
\begin{align*}
    \bar{x}_t &= \frac{\sum_{i: t_i = t} w_i x_i }{\sum_{i: t_i = t} w_i}, \quad t \in \{0,1\}.
\end{align*}
The weighted covariate variance is defined as 
\begin{align*}
    \sigma_t^2 = \frac{\sum_{i: t_i = t} w_i}{(\sum_{i: t_i = t} w_i)^2 - \sum_{i: t_i = t} w_i^2}\sum_{i: t_i = t} w_i(x_i - \bar{x}_t)^2.
\end{align*}

With the weighted covariate mean and variance, LSPS computes the standardized mean difference to assess covariate balance.

\section*{S2. Outcome Model for Survival Analysis}
In the empirical studies of Section \ref{sec:emp}, we use a Cox proportional hazards model (Cox, 1972) to estimate an unbiased hazard ratio (the outcome $Y$ is time to event). The Cox model is expressed by the hazard function denoted by $h(\tau)$. Within a subclass $s$, the subclass-specific hazard function $h^s(\tau)$ is estimated as,
\begin{align*}
    h^s(\tau) &= h^s_0(\tau)\exp(\zeta_s t),
\end{align*}
where $\tau$ is the survival time, $h^s_0(\tau)$ is the baseline hazard at time $\tau$, $t$ is the treatment, and $\exp(\zeta_s)$ is the subclass-specific hazard ratio of the treatment. This expression gives the hazard function at time $\tau$ for subjects with treatment $t$ in subclass $s$. 

The parameters in the Cox model are estimated by optimizing the likelihood
\begin{align*}
    L(\zeta_s) 
    &= \prod_{i:C_i = 1 \cap S_i = s}\frac{\exp(\zeta_s t_i)}{\sum_{j:Y_j\geq Y_i} \exp(\zeta_s t_j)},
\end{align*}
where $C_i = 1$ indicates the occurrence of the outcome.

The hazard ratio can be obtained by reweighting $\exp(\zeta_s)$ by the size of the subclass. However, in practice, due to within-subclass zero counts and finite machine precision, the hazard ratio is estimated by optimizing the Cox partial conditional across subclasses as in the Cyclops R package \citep{suchard2013}.

\section*{S.3 Proof of Theorem 2}
We prove Theorem 2, which states that under certain monotonicity assumptions about the effect of the unmeasured confounder on the treatment and on the outcome, the average treatment effect adjusting for the all measured covariates lies between the true effect and the effect adjusting for only the measured confounders.

Let $T$ be a binary treatment, $Y$ be an outcome, $U$ be an ordinal unmeasured confounder, $U \in \{1, \dots, K\}$, and $\mbX$ be some measured covariates. Furthermore, let $\mbX_C$ be the measured confounders, and $\mbX^\complement_C$ be the complement set of the measured confounders, $\mbX^\complement_C = \{\mbx: \mbx \in \mbX : \mbx \not\in \mbX_C\}$. We assume that the measured covariates $\mbX^\complement_C$ form a noisy measurement of $U$. We denote this noisy measurement as $U'$. The relationship between the true unmeasured confounder $U$ and the noisy measurement $U'$, is established by the following misclassification probabilities,  $p_{ij} = p(U' = i \g U = j)$, $i, j \in \{1, \dots, K\}$. Then, $p(U' = i) = \sum_{j=1}^Kp_{ij}p(U=j)$. 

We assume that the misclassification probabilities of $U$ is nondifferential with respect to $T$, $Y$, and $\mbX$; that is, $p(U'=u' \g U=u, Y=y, T=t, \mbX=\mbx) = p(U'=u' \g U=u, Y=y', T=t', \mbX=\mbx')$ for all $y, y' \in \mathcal{Y}$, $t, t' \in \{0,1\}$, and $x, x' \in \mathcal{X}$. If $p_{ij} \leq p_{ik}$ and $p_{ji} \leq p_{ki}$ for $j < k < i$ , and $p_{il} \geq p_{im}$ and $p_{li} \geq p_{mi}$ for $i < l < m$, then we say that the misclassification probabilities are tapered.

The proof of this theorem relies on Lemma 1-4. We extend Lemma 1-4 in Ogburn and VanderWeele \citep{ogburn2013} to conditions where identifiability holds by conditioning on measured and unmeasured confounders. 

\begin{lemma}
If $\EE{X_C}{Y \g T=1} \geq \EE{X_C, U'}{Y \g T=1} \geq \E{Y(1)}$ and $\EE{X_C}{Y \g T=0} \leq \EE{X_C, U'}{Y \g T=0} \leq \E{Y(0)}$, or if  $\EE{X_C}{Y \g T=1} \leq \EE{X_C, U'}{Y \g T=1} \leq \E{Y(1)}$ and $\EE{X_C}{Y \g T=0} \geq \EE{X_C, U'}{Y \g T=0} \geq \E{Y(0)}$, then the treatment effect adjusting for all measured covariates falls between the true effect and the effect adjusting for the measured confounders alone.
\end{lemma}

\begin{lemma}
If $\E{Y \g T, X, U}$ and $\E{T \g X, U}$ are either both nonincreasing or both nondecreasing in $U$, then $\EE{X_C}{Y \g T=1} \geq \E{Y(1)}$ and $\EE{X_C}{Y \g T=0} \leq \E{Y(0)}$. If one of $\E{Y \g T, X, U}$ and $\E{T \g X, U}$ is nonincreasing and the other nondecreasing in $U$, then $\EE{X_C}{Y \g T=1} \leq \E{Y(1)}$ and $\EE{X_C}{Y \g T=0} \geq \E{Y(0)}$. 
\end{lemma}

\begin{lemma}
Suppose that $U$ is nondifferentially misclassified with respect to $T$ and $Y$. If $\E{Y \g T, X, U}$ and $\E{T \g X, U}$ are both nondecreasing or both nonincreasing in $U$, then $\EE{X_C, U'}{Y \g T=1} \geq \E{Y(1)}$ and $\EE{X_C, U'}{Y \g T=0} \geq \E{Y(0)}$. If one of $\EE{X_C, U'}{Y \g T, X, U}$ and $\E{T \g X, U}$ is nondecreasing and the other nonincreasing in $U$, then $\EE{X_C, U'}{Y \g T=1} \leq \E{Y(1)}$ and $\EE{X_C, U'}{Y \g T=0} \leq \E{Y(0)}$.
\end{lemma}

\begin{lemma} Suppose that $U$ is nondifferentially misclassified with respect to $T$ and $Y$ with tapered misclassification probabilities. If $\E{Y \g T, X, U}$ and $\E{T \g X, U}$ are both nondecreasing or both nonincreasing in $U$, then $\EE{X_C, U'}{Y \g T=1} \leq \EE{X_C}{Y \g T=1}$ and $\EE{X_C, U'}{Y \g T=0} \leq \EE{X_C}{Y \g T=0}$. If one of $\E{Y \g T, X, U}$ and $\E{T \g X, U}$ is nondecreasing and the other nonincreasing in $U$, then $\EE{X_C, U'}{Y \g T=1} \geq \EE{X_C}{Y \g T=1}$ and $\EE{X_C, U'}{Y \g T=0} \geq \EE{X_C}{Y \g T=0}$.
\end{lemma}

Lemma 2 establishes the relationship between the true expectations and expectations conditioning on the measured confounders. Lemma 3 establishes the relationship between expectations conditioning on all measured covariates and the true expectations. Lemma 4 establishes the relationship between expectations conditioning on all measured covariates and expectations conditioning on the measured confounders. Lemma 2-4 establish the conditions of Lemma 1, and Theorem 2 then follows.





\section*{References}
P. C. Austin, An Introduction to Propensity Score Methods for Reducing the Effects of Confounding in Observational Studies, Multivariate Behavioral Research 46(3) (2011) 399–424.

D. R. Cox, Regression Models and Life-Tables, JSTOR 34 (2) (1972) 187–220.

E. L. Ogburn, T. J. Vanderweele. Bias attenuation results for nondifferentially mismeasured ordinal and coarsened confounders, Biometrika 100(1) (2013) 241–248. 

P. R. Rosenbaum, D. B. Rubin, Reducing Bias in Observational Studies Using Subclassification on the Propensity Score, Journal of the American Statistical Association 79 (387) (1984) 516.

M. A. Suchard, S. E. Simpson, I. Zorych, P. Ryan, D. Madigan, Massive parallelization of serial inference algorithms for complex generalized linear models, ACM Transactions on Modeling and Computer Simulation 23 (2013) 1-17.

\end{document}